\documentclass{article}

\pdfoutput=1
\usepackage{graphicx}
\usepackage{arxiv}

\usepackage[utf8]{inputenc} 
\usepackage[T1]{fontenc}    
\usepackage{url}            
\usepackage{booktabs}       
\usepackage{amsfonts}       
\usepackage{nicefrac}       
\usepackage{microtype}      
\usepackage{lipsum}

\title{Practical guide of using Kendall's $\tau$ in the context of forecasting critical transitions}

\author{ Shiyang Chen \\
  Department of Mechanical Engineering\\
  University of Michigan, Ann Arbor\\
  \texttt{shychen@umich.edu} \\
   \And
 Amin Ghadami \\
    Department of Mechanical Engineering\\
  University of Michigan, Ann Arbor\\
  \texttt{aghadami@umich.edu} \\
  \AND
  Bogdan I. Epureanu\\
  Department of Mechanical Engineering\\
  University of Michigan, Ann Arbor\\
  \texttt{epureanu@umich.edu} \\
}

\begin{document}
\maketitle

\begin{abstract}
Recent studies demonstrate that trends in indicators extracted from measured time series can indicate approaching to an impending transition. Kendall's $\tau$ coefficient is often used to study the trend of statistics related to the critical slowing down phenomenon and other methods to forecast critical transitions. Because statistics are estimated from time series, the values of Kendall's $\tau$ are affected by parameters such as window size, sample rate and length of the time series, resulting in challenges and uncertainties in interpreting results. In this study, we examine the effects of different parameters on the distribution of the trend obtained from Kendall's $\tau$, and provide insights into how to choose these parameters. We also suggest the use of the non-parametric Mann-Kendall test to evaluate the significance of a Kendall's $\tau$ value. The non-parametric test is computationally much faster compared to the traditional parametric ARMA test. 
\end{abstract}


\section{Introduction}
{Complex systems might undergo abrupt transitions from one stable state to another \cite{scheffer2009critical,scheffer2012anticipating,drake2010early,ghadami2018model,chen2018forecasting,dakos2017identifying,chen2019eigenvalues}.
Such a transition is usually undesirable, leading to various types of stability issues and possible catastrophic consequences. Specifically, regime shifts in ecological systems have received growing attention as the cumulative human impact on the environment has increased the risk of ecological regime shifts \cite{ghadami2018model}. The prediction of critical transitions faces significant challenges because changes in the equilibrium state of the system are generally small prior to transitions. Recent studies demonstrate that several indicators based on changes in ecological time series can indicate that the system is approaching an impending transition \cite{dakos2012methods,kuehn2011mathematical,kefi2013early,scheffer2009early}.
These indicators, called early warning indicators, are statistical measures that reveal proximity to a tipping point developed based on the slowing down phenomenon~\cite{strogatz2014nonlinear}.  When a dynamical system approaches a tipping point, its dynamics becomes progressively slow, phenomenon known as slowing down. As a consequence of the slowing down phenomenon, an increase in the values of certain statistics, i.e. variance and lag-1 autocorrelation of stochastic fluctuations of the system, have been observed prior to critical transitions in numerous theoretical and experimental complex systems ~\cite{scheffer2012anticipating,drake2010early}.}

To probe for indications of critical slowing down prior to a transition, the trend of the extracted warning indicators is monitored as system parameters gradually change. An increasing (positive) trend in the values of early warning indicators over time is considered as a sign of approaching a transition. Such a trend, however, needs to be quantified to allow one to analyze the changes in the system dynamics. In addition, identifying the trend of early warning signals might not be trivial due to stochastic fluctuations in the reported values of early warning signals over time.  As a result, Kendall's $\tau$ coefficient is often used to quantify the trend of statistics related to the critical slowing down phenomenon~\cite{dakos2012methods,ghadami2020data}. Kendall's $\tau$ is a measure of the correlation between the rank order of the observed values and their order in time~\cite{kendall1948rank}. A positive Kendall's $\tau$ typically means a monotonic increase in the data. However, interpreting the values of Kendall's $\tau$ is challenging and requires careful considerations. First, there exist a probability distribution over the possible values of Kendall's $\tau$ corresponding to each measurement. Hence, a positive Kendall’s $\tau$ does not guarantee that the system is moving toward a transition unless its significance is confirmed \cite{ghadami2020data,lenton2012early,boettiger2012quantifying}. Even for a signal measured from a stationary system, one might obtain a positive value for Kendall’s $\tau$. In addition, the values of Kendall's $\tau$ are affected by parameters such as window size, sample rate and the length of the time series corresponding to data collection and analysis steps \cite{ghadami2020data,lenton2012early,boettiger2012quantifying}. Hence, one needs to examine the effects of different parameters on the distribution of the trend statistic Kendall’s $\tau$. Without such a study and detailed understanding of the statistical significance of the estimated Kendall’s $\tau$ from measured time series, it would be difficult to conclude if a detected warning signal is a false alarm or not.

A number of parametric and non-parametric methods have been proposed to understand the significance of Kendall's $\tau$ values obtained from time series~\cite{dakos2012methods,boettiger2012quantifying}. Parametric tests can be more powerful, but require more information about the system. In contrast, non-parametric trend tests, such as the Mann-Kendall test, require only that data be independent and tolerate outliers~\cite{hamed1998modified}. Here, we study the effect of different parameters on the distribution of Kendall's $\tau$ coefficients and compare the application of parametric and non-parametric tests in identifying the significance of Kendall's $\tau$ values obtained from ecological time series measurements. We discuss the benefits and drawbacks of each method in the context of Kendall's $\tau$ coefficient and provide a practical guide of using Kendall's $\tau$ coefficients as an indicator for critical transitions. Results of this study may improve the reliability of predictions made about the risk of critical transitions in complex systems based on early warning signals.

\section{Effects of data availability and data processing on Kendall's $\tau$ statistics}
In this section we use a simple example system to highlight how different parameters of the analysis affect the distribution of Kendall's $\tau$ for systems either facing or not facing a critical transition. Simulation data is obtained from the harvesting model~\cite{may1977thresholds}:
\begin{eqnarray}
\label{eq:harvest}
dx = \left(r x (1-\frac{x}{K})+c\frac{x^2}{x^2+1}\right)dt+\sigma dW,
\end{eqnarray}
where $x$ is the amount of biomass, $K$ is the carrying capacity, $r$ is the maximum growth rate, $c$ is the maximum grazing rate, and $\sigma$ is the standard deviation of the white noise $dW$. Values of the parameters (except for the bifurcation parameter $c$) are selected as $K=10$, $r=1$, and $\sigma=0.01$ .
To use Kendall's $\tau$ to detect critical slowing down, target statistics such as the variance, or the autocorrelation are first calculated using a moving window. This sequence of statistics is then used to calculate Kendall's $\tau$, which is further used to make a decision about the system. In this study, we used variance as the statistic.

For a sequence of independent and randomly ordered data, i.e. when there is no trend or serial correlation structure among the observations, the trend statistic Kendall's $\tau$ should tend to a normal distribution for a large number $n$ of observations in the sequence. The normal distribution has mean zero and variance given by
\begin{eqnarray}
\label{eq:var}
var(\tau) = \frac{2(2n+5)}{9n(n-1)}.
\end{eqnarray}

Generic early warning signals, such as the Kendall's $\tau$ of autocorrelation, however, are calculated using a sequence of statistics that are obtained from the time series using a moving window. Positive correlation among the observations increases the chance of obtaining a large Kendall's $\tau$, even in the absence of an actual trend. Therefore, the choice of parameters, such as window size and sample rate, affects the correlation in data, and thus affect the distribution of Kendall's $\tau$. 

\subsection{Window size}
The choice of window size has a large influence on the distribution of Kendall's $\tau$. This is because a positive serial correlation exists when two consecutive moving windows have an overlap. This correlation is even stronger as the size of the moving window increases. 

\begin{figure}[!htb]
\centering
\includegraphics[width=\linewidth]{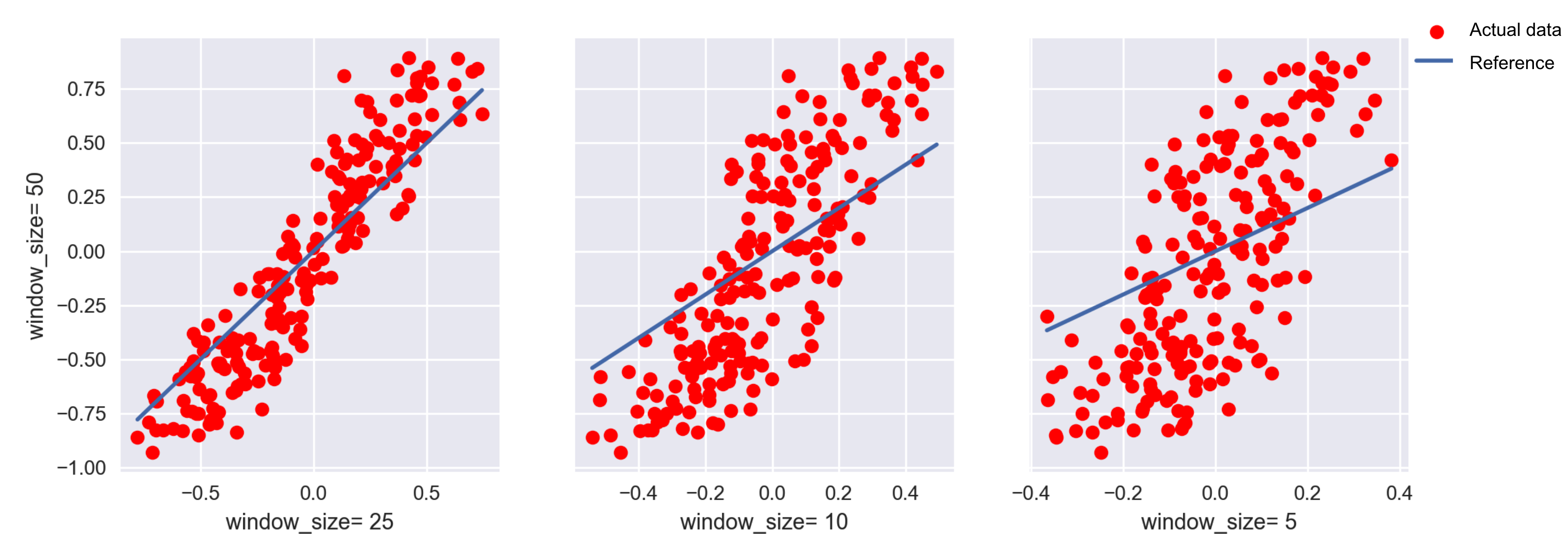}
\caption{ \label{fig:comp} Comparison between Kendall's $\tau$ calculated using a smaller window size and a large window size}
\end{figure}

To show this, we collected 400 time series from the harvesting model in Eq.~\ref{eq:harvest} with a fixed parameter value $c=-1.1$, and calculated Kendall's $\tau$ using a different window size for each time series. Figure~\ref{fig:comp} shows the relationship between Kendall's $\tau$ calculated using a smaller window size and a large window size. Each dot in the plot represents a result obtained from one time series calculated by solving Eq.~\ref{eq:harvest}. All time series have the same total duration of 1000. Results show that the curve takes an S shape as the difference between window sizes increases. This means that a large window size will inflate the value of Kendall's $\tau$ calculated from the same time series/data.

\begin{figure}[!htb]
\centering
\includegraphics[width=\linewidth]{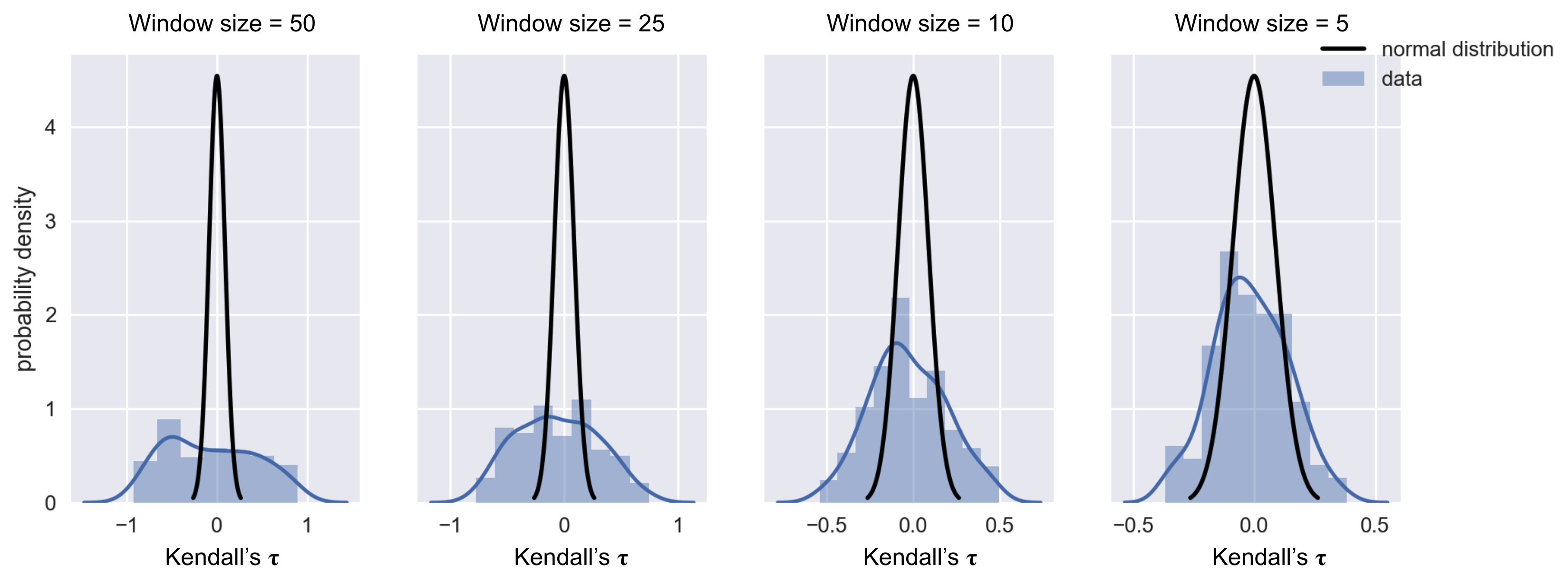}
\caption{ \label{fig:dist} Distribution of Kendall's $\tau$ calculated using different window sizes}
\end{figure}

The inflation of Kendall's $\tau$ for large window sizes can also be observed using the distribution plot shown in Fig.~\ref{fig:dist}. As the window size increases, the distribution of Kendall's $\tau$ becomes flatter, and farther away from the normal distribution with variance given by Eq.~\ref{eq:var}.

Therefore, the same Kendall's $\tau$ value has a completely different meaning with a different window size. A 90 \% percentile Kendall's $\tau$ value when the window size is 5 \% of the length of the time series is only 60 \% percentile Kendall's $\tau$ value when the window size is 50 \% of the length. Thus, merely calculating Kendall's $\tau$ values is not enough to decide the probability of a critical transition. A hypothesis test is necessary. A discussion about that is provided in Sec. 3. 

\subsection{Number of observations}

The number of observations in each of the data sets is also important. That is because all statistics will have a larger estimation error when only a limited number of observations are available. Moreover, it is harder to detrend the time series data when only a limited number of observations are available. Improper detrending may remove the important low frequency information, leaving behind only high frequency random noise. 

\begin{figure}[!htb]
\centering
\includegraphics[width=\linewidth]{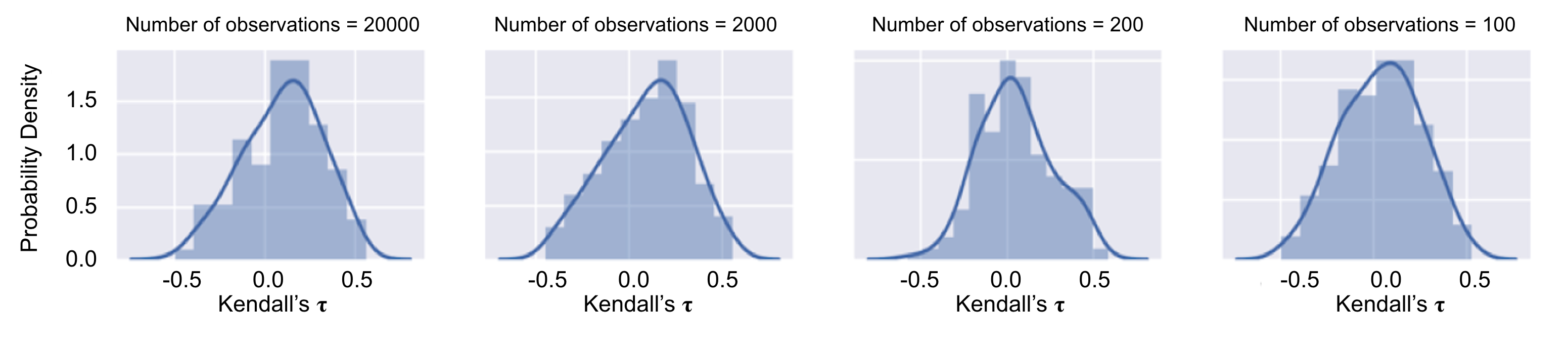}
\caption{ \label{fig:dist_sd} Distribution of Kendall's $\tau$ when 20000, 2000, 200 and 100 observations are used}
\end{figure}

To understand how the sampling rate can affect results, one time series data is obtained from the harvesting model in Eq.~\ref{eq:harvest}. The time series data is then down sampled to obtained another time series data with a smaller number of observations. The effect of the number of available data points can be observed in the distribution of the Kendall's $\tau$ when different numbers of observations are available. We again collected 400 time series of equal time length from the harvesting model (Eq.~\ref{eq:harvest}), with parameter $c$ continuously changing with time from $c=-0.6$ to $c=-1.1$. Each time series has 20,000 observations. These 400 time series are then down sampled to obtain time series with 2000, 200 and 100 observations. The resulting time series are then detrended to study the statistics around the system equilibrium, a standard procedure in the studies of early warning indicators of critical transitions~\cite{dakos2012methods}. The distribution of Kendall's $\tau$ are shown in Fig. \ref{fig:dist_sd}. When there are at least 2000 observations, the distribution of Kendall's $\tau$ is skewed toward the right, which is correct because the system is approaching the critical transition. However, when there are only 200 observations or fewer, the distribution becomes almost symmetric about zero, which is associated normally with random signals. Therefore, it is important to have enough observations in the data when Kendall's $\tau$ is used as an early warning signal, especially when the equilibrium of the system is changing as the system approaches the critical transition and detrending is necessary.

\section{Parametric and non-parametric tests to evaluate the significance of Kendall's $\tau$ values}
Because Kendall's $\tau$ is affected by several parameters, as discussed in the previous section, merely observing the values of Kendall's $\tau$ does not always reveal the desired information about the system. A number of tests have been proposed to understand the significance of a certain value of Kendall's $\tau$ in the literature. These approaches can be categorized as parametric and non-parametric tests. In this section, we review the parametric and non-parametric tests and compare them in the context of early warning signals. Particularly, we introduce the modified Mann-Kendall trend test as a potential and efficient method to evaluate the significance of Kendall's $\tau$ values of early warning signals where data availability is limited.

\subsection{Non-parametric Mann-Kendall trend test}

The non-parametric Mann-Kendall test is commonly employed to detect monotonic trends in time series. The null hypothesis for the traditional Mann-Kendall trend test, however, is that there is no trend or serial correlation structure among the observations. This hypothesis is not rigorously true for the series of statistics, such as the standard deviation, or the autocorrelation that are obtained from the time series using a moving window. As a result, using the standard Mann-Kendall test does not lead to reliable results in identifying the significance of the trend of the early warning signals~\cite{boettiger2012quantifying}.

Hamed and Rao~\cite{hamed1998modified} point out that a modified Mann-Kendall trend test can be used to study data with a serial correlation structure. In the modified test, the null hypothesis is that there is no trend in the data, but there can be autocorrelation, addressing the challenge existing in the studies of early warning signals. If the null hypothesis is true, Kendall's $\tau$ should follow a normal distribution with mean 0, and variance given by ~\cite{hamed1998modified}
\begin{eqnarray}
\label{eq:var2}
var(\tau) = \frac{2(2n+5)}{9n(n-1)}\left(1+\frac{2}{n(n-1)(n-2)}\sum_{i=1}^{n-1}(n-i)(n-i-1)(n-i-2)\rho_S(i)\right),
\end{eqnarray}
where $\rho_S(i)$ is the autocorrelation of the ranks of the observation, and $n$ is the number of observations. 

\begin{figure}[!htb]
\centering
\includegraphics[width=\linewidth]{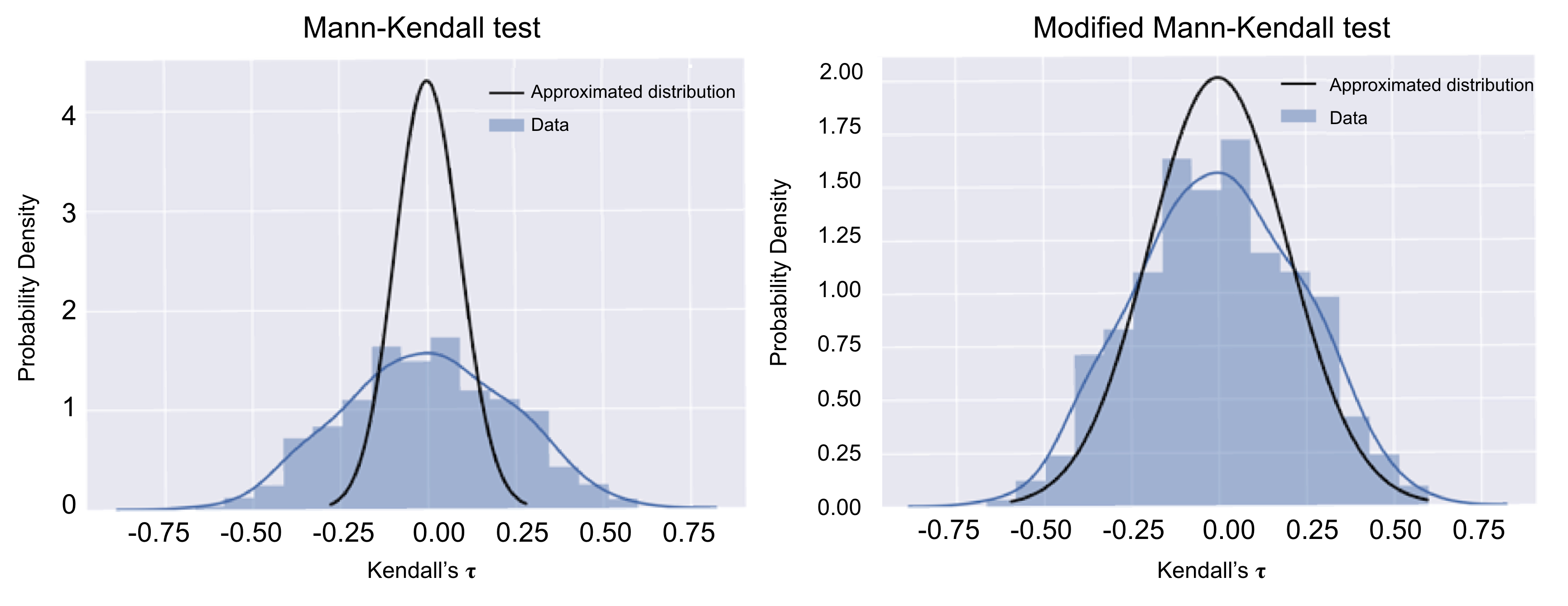}
\caption{ \label{fig:modified}Comparison between the distribution of Kendall's $\tau$ in data and the Mann-Kendall test (left), and comparison between the distribution of Kendall's $\tau$ in data and the modified Mann-Kendall test (right)}
\end{figure}

To better understand this, we compared the distribution of Kendall's $\tau$ to the normal distribution with variance calculated using Eq.~\ref{eq:var} and the modified distribution with variance calculated using Eq.~\ref{eq:var2}. In this example, we use 200 distinct time series generated using the harvesting model  (Eq. 1). For each time series, we calculated the variance as the early warning signal using a moving window. The window size and number of observations are selected as 100 and 12000, respectively. The values of Kendall's $\tau$ are then calculated for each set of obtained warning signals. The distribution of Kendall's $\tau$ obtained using the generated time series is shown in Fig.~\ref{fig:modified} revealing that the real distribution is much flatter than the normal distribution due to the positive correlation in data. Next, a single time series is used to calculate the varience of the normal distribution obtained by the modified Mann-Kendall test (i.e., Eq. (3)). Results shown in Fig~\ref{fig:modified} show that the modified distribution is much closer to the real distribution. 

In practice, one may typically only have one time series. Therefore, the real distribution is not available. In this case, we can use the available time series to calculate the modified distribution of Kendall's $\tau$, and use the distribution to calculate the percentile of the obtained Kendall's $\tau$ value.

\subsection{Parametric tests}

Parametric tests~\cite{dakos2012methods,boettiger2012quantifying} are a group of tests that have been proposed to study the significance of Kendall's $\tau$ values. These methods use a general model to fit the data, and generate artificial data using the model to understand how significant the trend statistic value is. Dakos et al.~\cite{dakos2012methods} proposed to fit an auto-regressive, moving-average model (ARMA) using the residual data after detrending. This test is designed to show that the data cannot come from a linear stationary process if a large Kendall's $\tau$ value is obtained from it. When this test gives a p-value as low as 0.1\%, that does not mean that the probability of critical transition is as high as 99.9 \%. It means that the probability that this time series data is generated using a linear stationary model is as low as 0.1 \%. Boettiger and Hastings~\cite{boettiger2012quantifying} proposed to fit two nonlinear models that both have a normal form for the saddle-node bifurcation. The difference between these two models is that one of them has a fixed bifurcation parameter, while the other has a changing parameter. The distributions of the test statistics generated from these two models compared to determine if these two models are statistically different and to explore why one of them better describes the data. 

\begin{figure}[!htb]
\centering
\includegraphics[width=0.8\linewidth]{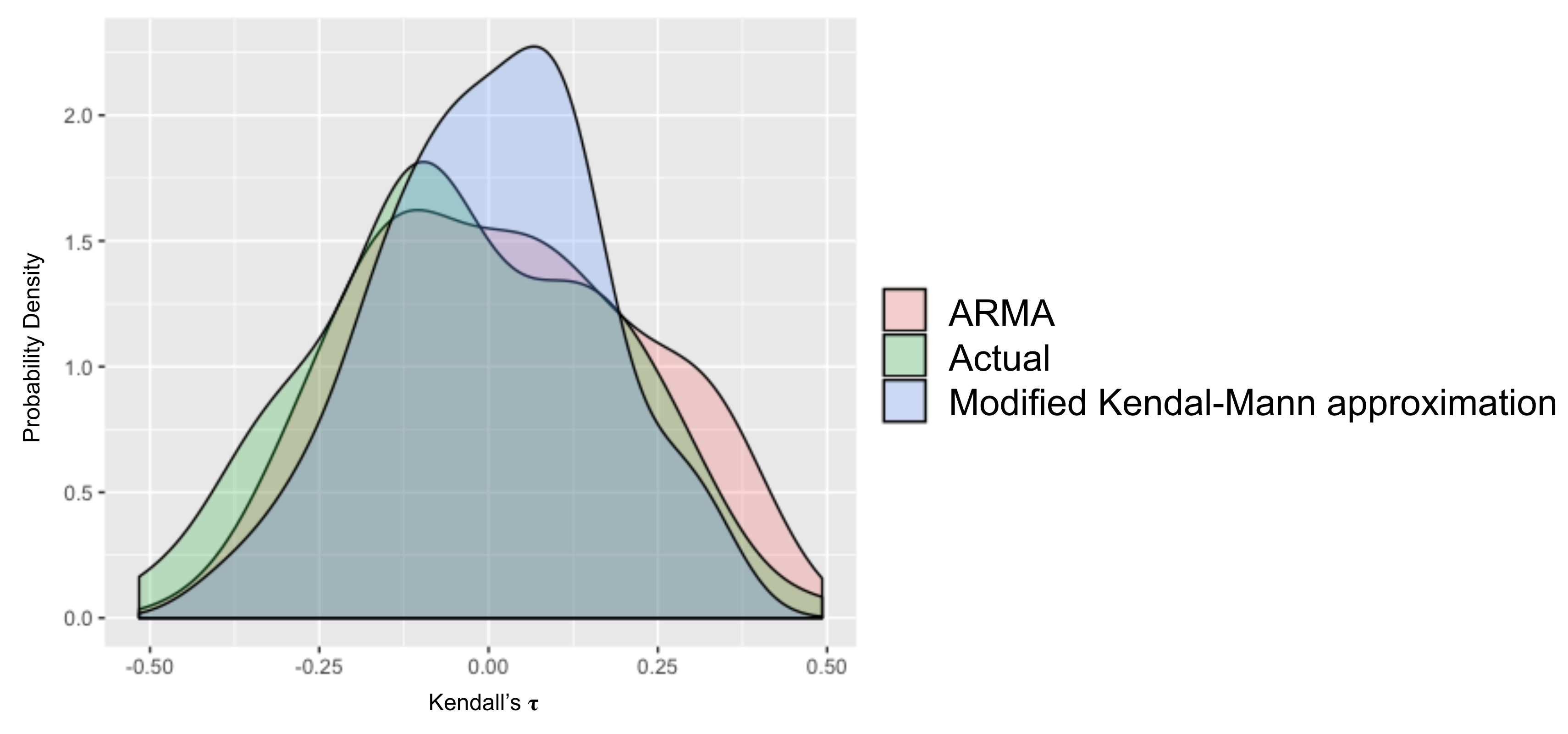}
\caption{ \label{fig:dist_kendall} Distribution of Kendall's $\tau$ calculated using the harvesting model, the fitted ARMA model, and the modified Mann-Kendall test}
\end{figure}

Here, we consider an ARMA model as an example of parametric method and evaluate its performance in approximating the Kendall's $\tau$ distribution obtained from the harvesting model (Eq.~\ref{eq:harvest}). Similar to the previous example, 200 distinct time series are first generated using the (harvesting) model. Next, we use one of the time series to fit an ARMA model, and generate another distinct 200 time series using the fitted ARMA model. The distribution of Kendall's $\tau$ calculated from the initially generated time series and all the ARMA time series are shown in Fig.~\ref{fig:dist_kendall}. Results show that the distribution approximated by the ARMA model are close to the reference distribution directly obtained by the time series of the harvesting model. 

For this example, we compared the results of the performed parametric test using an ARMA model and the non-parametric modified Mann-Kendall test.  Using the same single time series selected to generate ARMA results, the non-parametric modified Mann-Kendall distribution was approximated and plotted on the top of the other distributions in Fig.~\ref{fig:dist_kendall}. Figure~\ref{fig:dist_kendall} shows that all three distributions are close to each other. Therefore, both the parametric ARMA test and the non-parametric modified Mann-Kendall can accurately approximate the distribution of Kendall's $\tau$ calculated from times series data of low-dimensional systems with Gaussian noise. The benefit of the non-parametric Mann-Kendall test is that the distribution of Kendall's $\tau$ can be estimated directly from a single time series, and thus no further simulation or measurements are required in contrast to the parametric tests. As a result, the non-parametric modified Mann-Kendall test is much faster computationally than the parametric ARMA test.
\section{Discussion and Conclusions}
Kendall's $\tau$ is often used to quantify the trend of statistics related to the critical slowing down phenomenon and detect if the system is at risk of an upcoming transition.  Due to the probabilistic nature of the Kendall's $\tau$ values, however, making any conclusion about the risk of impending transitions based on these values requires a detailed understanding of the statistical significance of the estimated Kendall’s $\tau$ from time series. 

In this study, we examined how different analysis parameters can affect the value of Kendall's $\tau$ obtained from the same system, and we demonstrated why a significance test on the estimated Kendall's $\tau$ coefficient is necessary. We summarized and compared selected parametric and non-parametric tests to evaluate the significance of Kendall's $\tau$. Particularly, we proposed to use the non-parametric Mann-Kendall test as an efficient test to assess the reliability of the approximated  values. It was demonstrated that both the parametric and non-parametric tests yield similar and valid results for a low-dimensional system with Gaussian stochastic excitation. The benefit of the non-parametric Mann-Kendall test, however, is that the distribution of Kendall's $\tau$ can be estimated directly from the time series, and thus no further data or simulations are required, and the computation is much faster than the parametric tests. 

Based on this analysis, we suggest a guideline to consider when using Kendall's $\tau$ to study a system subject to critical slowing down. First, a large window size can inflate the value of Kendall's $\tau$ compared to a small window size. We encourage the use of smaller window sizes when there is a large enough amount of data.
Second,  we demonstrated that the values of Kendall's $\tau$ are sensitive to the number of available observations of the system. For a smaller number of available data, the probability of estimating random and irrelevant values of Kendall's $\tau$ is higher. Third, we propose that the significance of the obtained Kendall's $\tau$ values should be studied. Such a study can be performed using either the non-parametric Mann-Kendall test or parametric tests, but the non-parametric Mann-Kendall test was shown to be a more computationally efficient approach.

\section*{Acknowledgement}
This research was supported by the National Institute of General Medical Sciences of the National Institutes of Health under Award Number U01GM110744. The content is solely the responsibility of the authors and does not necessarily reflect the official views of the National Institutes of Health.
\newpage
\bibliographystyle{ieeetr}
\bibliography{references.bib}

\end{document}